\documentclass[aps,prl,amsfonts,amsmath,amssymb,reprint,twocolumn,superscriptaddress,showpacs,a4paper]{revtex4-1}

\usepackage[squaren]{SIunits}
\usepackage[usenames,dvipsnames]{color}
\usepackage[pdftex]{graphicx}
\usepackage{bm}
\usepackage{hyperref}
\usepackage{bbm}

\usepackage{color}
\definecolor{mygreen}{rgb}{0,0.5,0} 
\definecolor{myblue}{rgb}{0,0,0.75} 
\definecolor{myyellow}{rgb}{0.87,0.8,0.47} 
\definecolor{mymagenta}{cmyk}{0,1,0,0.12}

\newcommand{\rb}{$^{87}\mathrm{Rb}\ $}
\newcommand{\var}{\mathrm{var}}
\newcommand{\cov}{{\mathrm{cov}}}

\newcommand{\ave}[1]{\ensuremath{\langle#1\rangle}}
\newcommand{\norm}[1]{\ensuremath{\lvert#1\rvert}}

\newcommand{\NL}{N_{\rm L}}

\newcommand{\Jx}{\hat{J}_{\rm x}}
\newcommand{\Jy}{\hat{J}_{\rm y}}

\newcommand{\Sx}{\hat{S}_{\rm x}}
\newcommand{\Sy}{\hat{S}_{\rm y}}
\newcommand{\Sz}{\hat{S}_{\rm z}}

\newcommand{\bS}{\bm{S}}

\newcommand{\supin}{^{({\rm in})}}
\newcommand{\supout}{^{({\rm out})}}

\newcommand{\Fx}{\hat{F}_{\rm x}}
\newcommand{\Fy}{\hat{F}_{\rm y}}
\newcommand{\Fz}{\hat{F}_{\rm z}}

\newcommand{\rsup}[1]{^{({\rm #1})}}

\newcommand{\bF}{\hat{\bf F}}
\newcommand{\bff}{\hat{\bf f}}
\newcommand{\supi}{^{(i)}}

\begin{document}

\title{Generation of macroscopic singlet states in a cold atomic ensemble}

\newcommand{\ICFOAddress}{ICFO-Institut de Ciencies Fotoniques, Mediterranean Technology Park, 08860 Castelldefels (Barcelona), Spain}
\newcommand{\ICREAAddress}{ICREA -- Instituci\'{o} Catalana de Re{c}erca i Estudis Avan\c{c}ats, 08015 Barcelona, Spain}
\newcommand{\BilbaoAddress}{Department of Theoretical Physics, University of the Basque Country UPV/EHU, P.O. Box 644, E-48080 Bilbao, Spain}
\newcommand{\IkerbasqueAddress}{IKERBASQUE, Basque Foundation for Science, E-48011 Bilbao, Spain}
\newcommand{\HungaryAddress}{Wigner Research Centre for Physics, Hungarian Academy of Sciences, P.O. Box 49, H-1525 Budapest, Hungary}

\author{N. Behbood}
\affiliation{\ICFOAddress}
\email[]{naeimeh.behbood@icfo.es}

\author{F. Martin Ciurana}
\affiliation{\ICFOAddress}

\author{G. Colangelo}
\affiliation{\ICFOAddress}

\author{M. Napolitano}
\affiliation{\ICFOAddress}

\author{G\'{e}za T\'{o}th}
\affiliation{\BilbaoAddress}
\affiliation{\IkerbasqueAddress}
\affiliation{\HungaryAddress}

\author{R.J. Sewell}
\affiliation{\ICFOAddress}

\author{M.W.~Mitchell}
\affiliation{\ICFOAddress}
\affiliation{\ICREAAddress}

\date{\today}

\begin{abstract}
We report the generation of a macroscopic singlet state in a cold atomic sample via quantum non-demolition (QND) measurement induced spin squeezing. 
We observe 3 dB of spin squeezing and detect entanglement with $5\sigma$ statistical significance using a generalized spin squeezing inequality.  The degree of squeezing implies at least 50\% of the atoms have formed singlets.
\end{abstract}


\maketitle

Generating and detecting large-scale spin entanglement in many-body quantum systems is of fundamental interest~\cite{Lewenstein2007,LewensteinBOOK2012} and motivates many experiments with cold atoms~\cite{TrotskyPRL2010,SimonN2011,Nascimbene2012a,GreifPRL2011,GreifS2013} and ions~\cite{Islam2013a}.
For example, macroscopic singlet states appear as ground states of many fundamental spin models~\cite{AndersonS1987,BalentsN2010}, and even in quantum gravity calculations of black hole entropy~\cite{Livine2005a}.
Here we report the production of a macroscopic spin singlet (MSS) in an atomic system using collective quantum non-demolition (QND) measurement~\cite{KoschorreckPRL2010a,KoschorreckPRL2010b,SewellNP2013} as a global entanglement generator.

QND measurement is a well-established technique for generating conditional spin squeezing in polarized  atomic samples~\cite{Kuzmich1998,AppelPNAS2009,Takano2009,Schleier2010,Leroux2010a,Chen2011,SewellPRL2012}, where the state-of-the-art is 10 dB of squeezing in a cavity-enhanced measurement~\cite{Bohnet2013}.
In our experiment we apply QND measurement techniques to an unpolarized sample.
The QND measurement first generates large-scale atom-light entanglement by passing a macroscopic optical pulse through the entire ensemble.
The optical pulse is then measured, transferring the entanglement onto the atoms and leaving them in an entangled state~\cite{TothNJP2010}.  
Subsequent measurements on the ensemble confirm the presence of a MSS with a singlet fraction of approximately one half.
Our techniques are closely related to proposals for using QND measurement to detect~\cite{EckertNP2008,EckertPRL2007} and generate~\cite{HaukePRA2013} long-range correlations in quantum lattice gases and spinor condensates.

A MSS has a collective spin $\bF=0$, where ${\bF}\equiv\sum_{i}{\bff}^{(i)}$ and ${\bff}\supi$ is the spin of the $i$'th atom.
This implies that fluctuations in the collective spin vanish, i.e. $\Delta\bF=0$, suggesting that we can both produce and detect a macroscopic singlet via QND measurement induced spin squeezing~\cite{TothNJP2010, HaukePRA2013}.
Indeed, it has been shown that a macroscopic spin singlet can be detected via the generalized spin squeezing parameter
\begin{align}  
	\xi^2 = \frac{\sum_k(\Delta \hat{F}_k)^2}{f N_A}
\label{eq:spinSqueezingCriterion}
\end{align}
where $\xi^2<1$ indicates spin squeezing in the sense of
noise properties not producible by any separable state,
and thus  detects entanglement among the atoms~\cite{Toth2004a,Toth2007,Toth2009,TothNJP2010,Vitagliano2011a,Vitagliano2014a}.
The standard quantum limit (SQL) for unpolarized atoms is set by $\xi^2=1$, i.e. $\sum_k(\Delta \hat{F}_k)^2=f N_A$.
The number of atoms that are at least pairwise entangled in such a squeezed state is lower-bounded by $(1-\xi^2) N_A$~\cite{TothNJP2010}.
In the limit $\xi^2\rightarrow0$, the macroscopic many-body state is a true spin singlet.
Another criterion for detecting entanglement in non-polarized states has recently been applied to Dicke-like spin states~\cite{Bernd2014}. 
Our results complement recent work with quantum lattice gases~\cite{TrotskyPRL2010,Nascimbene2012a,GreifS2013}, and are analogous to the generation of macroscopic singlet Bell states with optical fields~\cite{Iskhakov2011,Timur2011}.

Since the collective spin obeys spin uncertainty relations $(\Delta \hat{F}_i)^2 (\Delta \hat{F}_j)^2 \ge |\ave{\hat{F}_k}|^2/4$ (we take $\hbar=1$ throughout), squeezing all three spin components requires maintaining an unpolarized atomic sample with $\ave{\hat{F}_k}\simeq0$.
Our experiment starts from a thermal spin state (TSS), i.e. a completely mixed state described by a density matrix $R=\rho^{\otimes N_A}$, where $\rho=\tfrac{1}{3}\mathbbm{1}_{3\times3}$ and $\mathbbm{1}_{3\times3}$ is the identity matrix.
This state has $\ave{\hat{F}_k}=0$ and $(\Delta\hat{F}_k)^2=(2/3) N_A$.
It is symmetric under exchange of atoms, and mixed at the level of each atom.

We probe the atoms via paramagnetic Faraday rotation using pulses of near-resonant propagating along the trap axis to give a high-sensitivity measurement of $\Fz$.
The optical pulses are described by Stokes operators ${\bS}$, which obey $[\Sx,\Sy]=i \Sz$ and cyclic permutations.  
The input pulses are fully $\Sx$-polarized, i.e. with $\ave{\Sx} = \NL/2$, where $\NL$ is the number of photons in the pulse.
During a measurement pulse, the atoms and light interact via an effective hamiltonian~\cite{Supp}
\begin{align}
\tau \hat{H}_{\rm eff} =  G_1 \Sz\Fz
\label{eq:effHamiltonian}
\end{align}
where $G_1$ is a coupling constant describing the vector lights shift and $\tau$ is the pulse duration~\cite{Echaniz2008,Colangelo2013a}.
Eq.~(\ref{eq:effHamiltonian}) describes a QND measurement of $\Fz$, i.e., a measurement with no back-action on $\Fz$.
We detect the output
\begin{align}
\Sy\supout=  \Sy\supin + G_1 \Sx\supin\Fz\supin
\label{eq:SyOut}
\end{align}
which leads to measurement-induced conditional spin squeezing of the $\Fz$ component by a factor $1/(1+\zeta)$, where $\zeta=\tfrac{2}{3}G_1^2N_LN_A$ is the signal-to-noise ratio (SNR) of the measurement~\cite{Hammerer2004}.

To measure and squeeze the remaining spin components, we follow a stroboscopic probing strategy described in Refs.~\cite{Behbood2013VB, Behbood2013SC}.
We apply a magnetic field along the [1,1,1] direction so that the collective atomic spin rotates $\Fz\rightarrow\Fx\rightarrow\Fy$ during one Larmor precession cycle.
We then time our probe pulses to probe the atoms at $T_L/3$ intervals, allowing us to measure all three components of the collective spin in one Larmor period.
Note that the probe duration $\tau\ll T_L$, so that we can neglect the rotation of the atomic spin during a probe pulse.

This measurement procedure respects the exchange symmetry of the input TSS, and generates correlations among pairs of atoms independent of the distance between them, leading to large-scale entanglement of the atomic spins.
The resulting state has $(1-\xi^2) N_A$ spins entangled in a MSS, and $\xi^2 N_A$ spin excitations (spinons).

\begin{figure}[t!]
\centering
\includegraphics[width=\columnwidth]{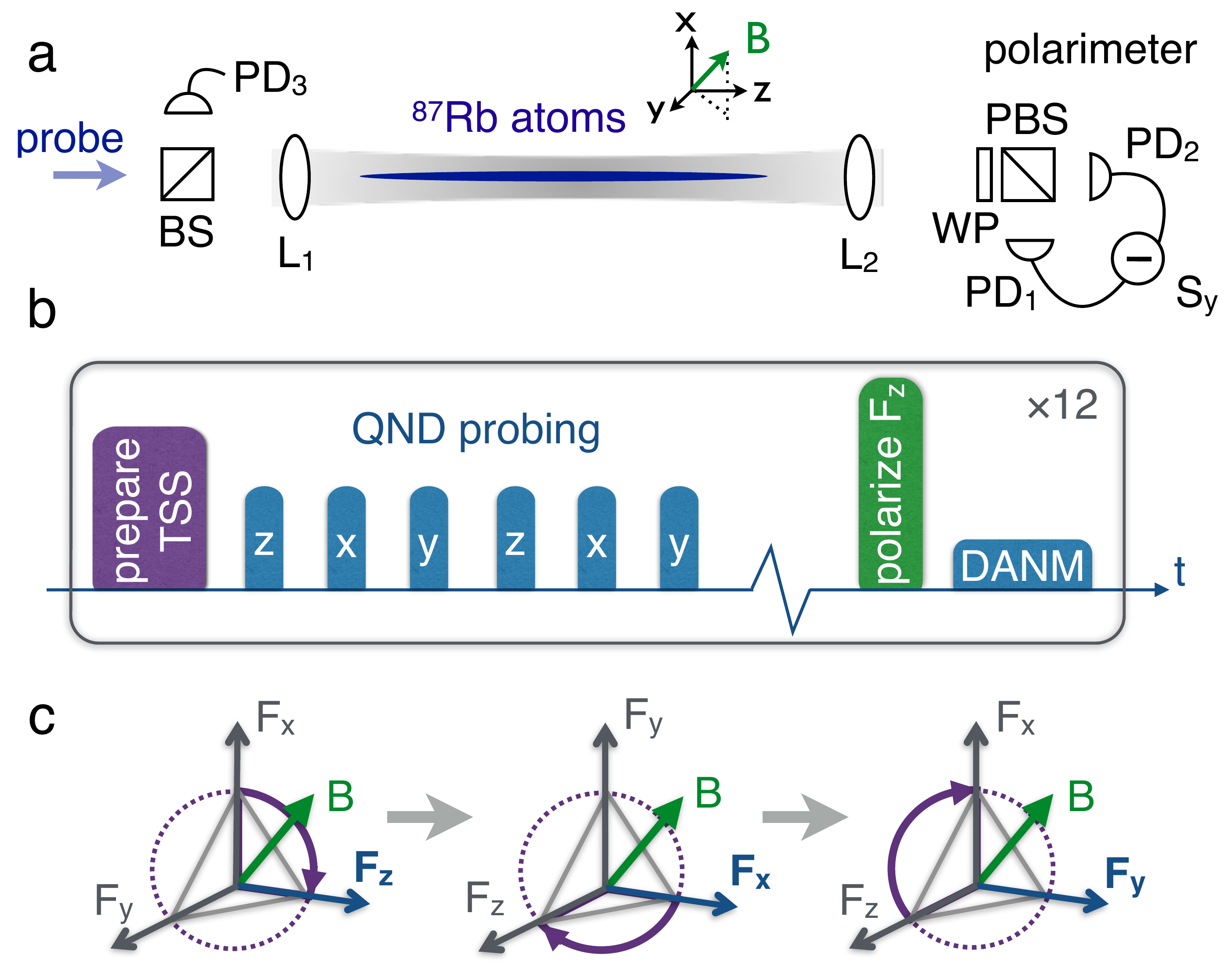}
\caption{(Color online) (a) Experimental geometry. 
Near-resonant probe pulses pass through a cold cloud of \rb atoms and experience a Faraday rotation by an angle proportional to the on-axis collective spin $\Fz$ . 
The pulses are initially polarized with maximal Stokes operator $\Sx$ recorded on reference detector (PD3). 
Rotation toward $\Sy$ is detected by a balanced polarimeter consisting of a wave plate (WP), polarizing beam splitter (PBS), and photodiodes (PD1,2). 
(b) Pulse sequence: A first QND measurement measures the $\Fz$  angular momentum component of the input atomic state, and the second and third QND measurements in 1/3 and 2/3 of Larmor precession cycles measure $\Fy$ and $\Fx$ respectively. 
(c) $\bF$ precesses about a magnetic field ({\bf B}) along the direction [111] making all components accessible to measurement via stroboscopic probing. 
\label{fig:setup}}
\end{figure}

\begin{figure}
\centering
\includegraphics[width=0.9\textwidth]{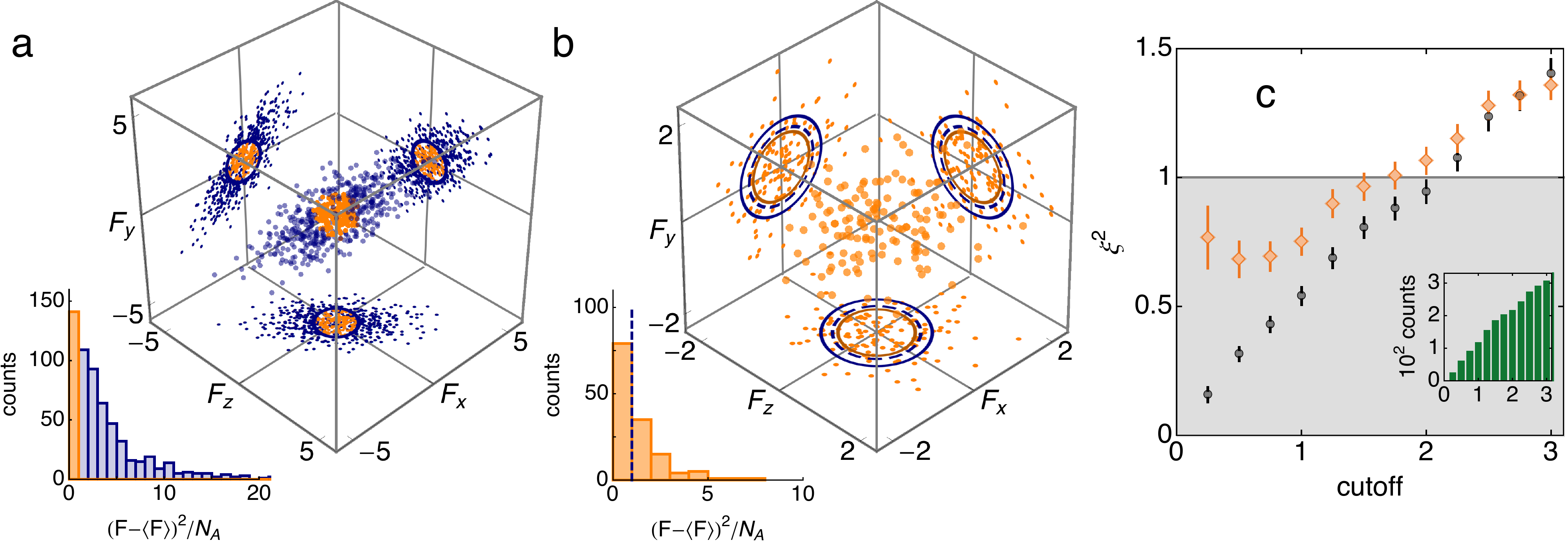}
\caption{(Color online) 
Selection of a macroscopic spin singlet.
From the initial spin distribution (blue data in figure (a)), we select data with $| \bF^{(1)} - \ave{\bF^{(1)}}|^2/N_A < C $ (orange data in figure (a)), where $C$ is a chosen cutoff parameter.
We then analyze the second QND measurement $\bF^{(2)}$ of the same data (orange data in figure (b)) to detect spin squeezing and entanglement.
We illustrate this with data from a sample with $N_A=1.1\times10^6$ atoms and $C=1$.
Axes in (a) \& (b) have units of $10^3$ spins.
In (a), the solid blue circle has a radius $\sqrt{C N_A}$.
In (b), the solid blue circle has a radius equal to the $1\sigma$ Gaussian RMS of an input ideal TSS with $\xi^2=2$, including detection noise, and the dashed blue circle the same for a state with $\xi^2=1$.
The solid orange circles in (b) indicates the measured $1\sigma$ Gaussian RMS of the selected data.
In the insets we plot a histogram of the first and second measurements.
The selected data are plotted in orange, and the dashed blue line in (b) indicates the cutoff.
(c) Spin squeezing parameter $\xi^2$ (orange diamonds) calculated from the second QND measurement of the selected data as a function of the cutoff parameter $C$.
The shaded region indicates $\xi^2<1$, i.e. spin squeezing according to the criterion given in Eq.~(\ref{eq:spinSqueezingCriterion}).
For reference, the same parameter calculated from the first QND measurement is also plotted (black circles).
Inset: number of selected data points included as a function of the cutoff parameter.
\label{fig:postSelection}}
\end{figure}

Our experimental apparatus, illustrated in Fig.~\ref{fig:setup}(a), is described in detail in Refs.~\cite{Kubasik2009}.
In each cycle of the experiment we trap up to $1.5 \times 10^6$ $^{87}$Rb atoms in a weakly focused single beam optical dipole trap.
The atoms are laser-cooled to a temperature of 20 $\mu$K, and optically pumped into the $f=1$ hyperfine ground state.
A shot-noise-limited balanced polarimeter detects $\Sy ^{\rm (out)}$ while a reference detector before the atoms measures $\Sx^{\rm (in)}$.
The trap geometry produces a large atom-light interaction for light pulses propagating along the axis of the trap, quantified by the effective optical depth $d_0=(\sigma_0/A)N_A$, where $\sigma_0=\lambda^2/\pi$ and $A=2.7\times10^{-9}$ m is the effective atom-light interaction area~\cite{Kubasik2009}, giving $d_0=69.5$ with $N_A=1.5\times10^6$ atoms.
We measure an atom-light coupling constant $G_1=9.0\pm0.1\times10^{-8}$ radians per spin~\cite{Supp}.
The measured sensitivity of the Faraday rotation probing is $\Delta F_z=515$ spins~\cite{KoschorreckPRL2010a}, allowing projection-noise-limited probing of an input TSS with $N_A>1.75\times10^5$ atoms.

The measurement sequence is illustrated in Figs.~\ref{fig:setup}(b),(c).
For each measurement, the atoms are initially prepared in a TSS via repeated optical pumping of the atoms between $f=1$ and $f=2$, as described in Ref.~\cite{KoschorreckPRL2010a}.
We then probe the atomic spins using a train of $\tau=1$ $\mu$s long pulses of linearly polarized light, detuned by $700$ MHz to red of the $f=1\rightarrow f'=0$ transition of the $D_2$ line.
Each pulse contains on average $N_L=  2.8 \times 10^8$ photons.
To access also $\Fx$ and $\Fy$, we apply a magnetic field with a magnitude $B= 16.9\pm0.1$ mG along the direction [111].
The atomic spins precess around this applied field with a Larmor period of $T_L = 85$ $\mu\mathrm{s}\gg\tau$, and we probe the atoms at $T_L/3 = 28.3$ $\mu\mathrm{s}$ intervals for two Larmor periods, allowing us to analyze the statistics of repeated QND measurements of the collective spin.

After the QND probing, the number of atoms $N_A$ is quantified via dispersive atom number measurement (DANM)~\cite{KoschorreckPRL2010a,KoschorreckPRL2010b} by applying a bias field $B_z=100$ mG and optically pumping the atoms into $\left| f=1, m_f=1 \right>$ with circularly-polarized light propagating along the trap axis resonant with the $f=1\rightarrow f'=1$ transition, and then probing with the Faraday rotation probe.

The sequence of state-preparation, stroboscopic probing and DANM is repeated 12 times per trap loading cycle. 
In each sequence $\sim15\%$ of the atoms are lost, mainly during the state-preparation, so that different values of $N_A$ are sampled during each loading cycle.
At the end of each cycle the measurement is repeated without atoms in the trap.
The loading cycle is repeated 602 times to gather statistics.

To detect the MSS, we make two successive measurements of the collective spin vector $\bF$ for each state preparation.
The first measurements give us a record of the input spin-distribution (blue points in Fig.~\ref{fig:postSelection}(a)).
The spread of these data includes contributions from technical noise in the atomic state preparation, and read-out noise in the detection system.
We select from the first measurements the events near the mean (orange points in Fig.~\ref{fig:postSelection}(a)), i.e. a low-dispersion subset of our data~\cite{Fukuhara2013a}.
The second measurement of these selected events, shown in Fig.~\ref{fig:postSelection}(b), is analyzed to determine if the selected subset satisfies the criterion for a MSS.

The selection procedure is illustrated in Figs.~\ref{fig:postSelection}(a) \& (b).
We select data from the first QND measurement of the collective spin vector using the criterion $| \bF - \ave{\bF}|^2/N_A < C $, where $C$ is a chosen cutoff parameter.
We calculate $\xi^2=\widetilde{\mathcal{V}}_2/(fN_A)$  from the second QND measurement, where  $\widetilde{\mathcal{V}}_2$ is the total variance after subtraction the read-out noise, $\widetilde{\mathcal{V}}_2\equiv\mathcal{V}_2-\mathcal{V}_0$.
Here $\mathcal{V}_2\equiv\mathrm{Tr}(\Gamma_2)$, where $\Gamma_2$ is the covariance matrix corresponding to the second QND measurement, and the read-out noise $\mathcal{V}_0\equiv\mathrm{Tr}(\Gamma_{0})$ is quantified by repeating the measurement without atoms in the trap and calculating the corresponding covariance matrix $\Gamma_{0}$.
For this experiment $\mathcal{V}_0= 9.2 \pm 0.3 \times 10^5$ spins$^2$.
This selection procedure is a form of measurement-induced spin squeezing~\cite{SewellPRL2012}, verified by the second QND measurement.
In Fig.~\ref{fig:postSelection}(c) we show $\xi^2$, computed on the second measurements of the selected events, as a function of the cutoff parameter $C$ for data from a sample with $N_A=1.1\times10^6$.
With a cutoff $C=0.75$ we measure $\xi^2=0.69\pm0.05$, detecting entanglement with $5\sigma$ significance.

We cross-check our results by repeating the experiment under near-identical conditions and analyzing the conditional covariance between successive vector spin measurements.
This allows us to deterministically prepare a MSS without filtering our data.
For these measurements the applied magnetic field had a magnitude $B= 15.9$ mG, giving a Larmor period of $T_L = 90\pm3$ $\mu$s, and we repeated the experiment 155 times.

\begin{figure}[b]
\centering
\includegraphics[width=\columnwidth]{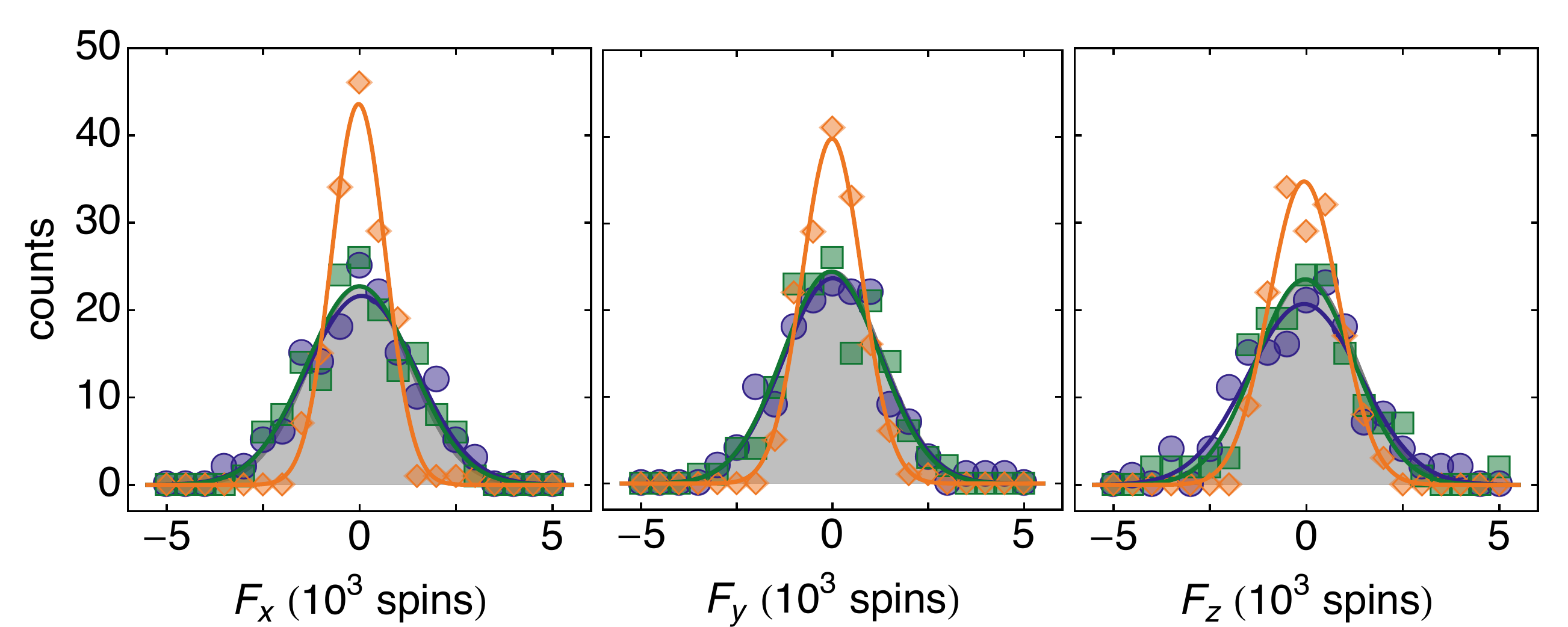}
\caption{Individual spin measurements.
Histograms of the measurements of each of the three spin components in the first round (blue circles) and second round (green squares) of the stroboscopic probe.
We also show the conditional spin distribution $F_k^{(2)}-\chi F_k^{(1)}$ (orange diamonds), where $\chi\equiv\cov(F_k^{(1)},F_k^{(2)})/(\Delta F_k^{(1)})^2$ is the degree of correlation.
The gray shaded region indicates the expected distribution for an ideal input TSS, including detection noise.
For presentation purposes an offset (between 5 and $10\times10^3$ spins) has been subtracted from the data~\cite{Supp}.
\label{fig:spinMeasurements}}
\end{figure}

Correlations between successive measurements of the same spin component $\hat{F}_k$ allows us to predict the outcome of the second measurements $F_k^{(2)}$ with a reduced conditional uncertainty.
For a single parameter, the conditional variance is $\var(F_k^{(2)}|F_k^{(1)})\equiv\var(F_k^{(2)}-\chi F_k^{(2)})$, where the {correlation parameter} $\chi\equiv\cov(F_k^{(1)},F_k^{(2)})/\var(F_k^{(1)})$ minimizes the conditional variance~\cite{SewellPRL2012}.
This is illustrated  in Fig.~\ref{fig:spinMeasurements}.

This procedure is readily extended to the conditional covariance using standard multivariate statistics.
We calculate the total variance $\mathcal{V}_{1,2}\equiv \mathrm{Tr}(\Gamma_{1,2})$ of the QND measurements, where $(\Gamma_{\rm c})_{ij} \equiv \cov(\hat{F}_i\rsup{c},\hat{F}_j\rsup{c}) \equiv \tfrac{1}{2}\ave{\hat{F}_i\rsup{c} \hat{F}_j\rsup{c}+\hat{F}_j\rsup{c}\hat{F}_i\rsup{c}}-\ave{\hat{F}_i\rsup{c}}\ave{\hat{F}_j\rsup{c}}$.
Conditional noise reduction is quantified via $\mathcal{V}_{2|1}=\mathrm{Tr}(\Gamma_{2|1})$, i.e. the total variance of the conditional covariance matrix $\Gamma_{2|1}\equiv\Gamma_{2}-\Gamma_{2,1}\Gamma_{1}^{-1}\Gamma_{1,2}$ where $\Gamma_{1,2}\equiv\cov(\hat{F}_i^{(1)},\hat{F}_j^{(2)})$~\cite{Kendall1979}.
To estimate the atomic noise contribution we fit the polynomial $\mathcal{V}_\alpha(N_A)=\mathcal{V}_0+2 N_A + c N_A^2$ to the measured data for the two QND measurements and the conditional variance.
We then calculate $\widetilde{\mathcal{V}}_{\alpha}=\mathcal{V}_{\alpha}-\mathcal{V}_{0}$, subtracting the read-out noise from the measured total variances.

\begin{figure}
\centering
\includegraphics[width=0.8\columnwidth]{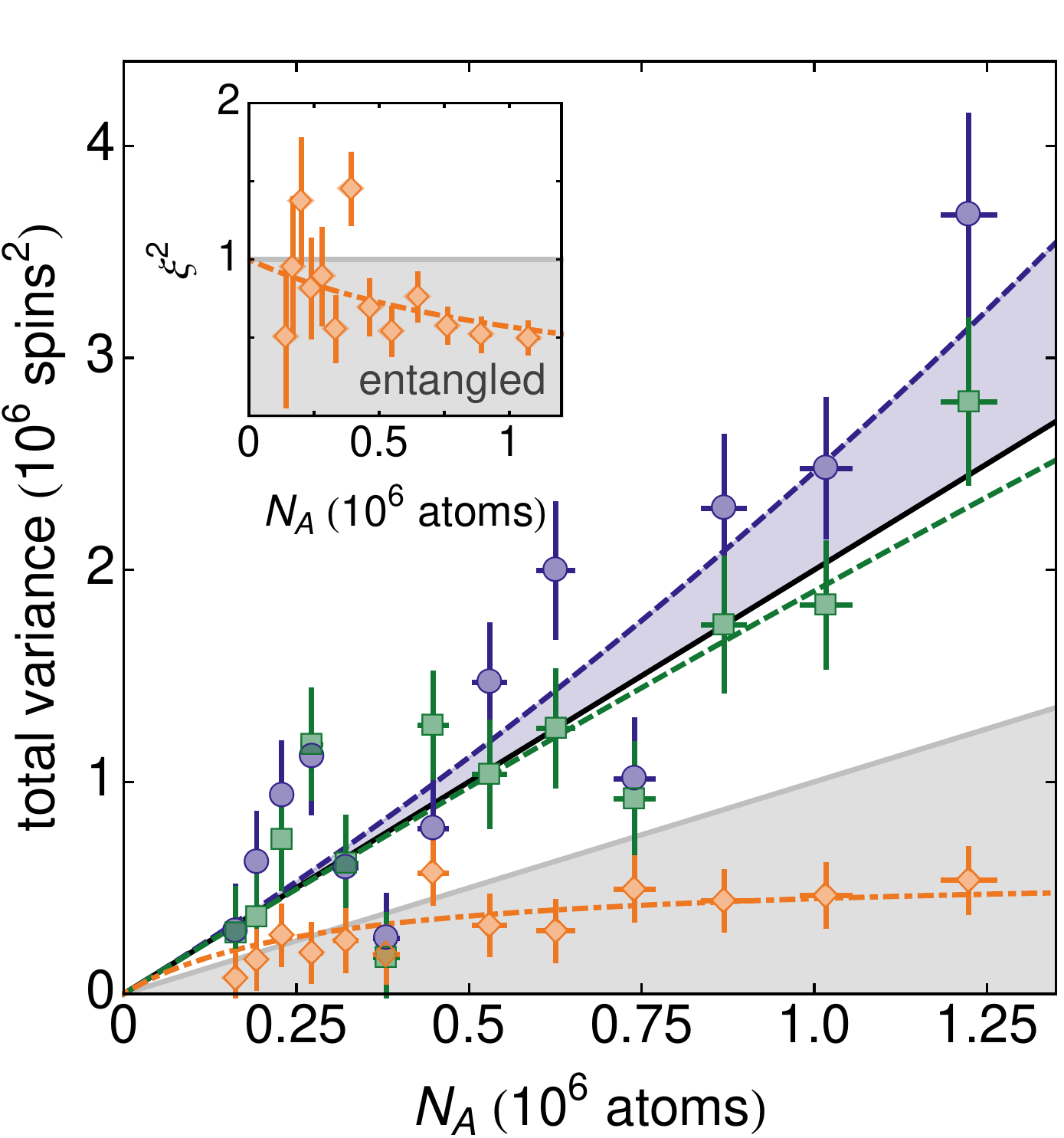}
\caption{(Color online) 
Noise scaling of total variance $\widetilde{\mathcal{V}}(N_A)$ of the first (blue circles) and second (green squares) QND measurement of the atomic spin distribution, and conditional variance $\widetilde{\mathcal{V}}_{2|1}$ (orange diamonds). 
Dashed lines are a quadratic fit, indicating the presence of technical noise in the input atomic state (blue shaded region).
Black line indicates the expected quantum noise $\widetilde{\mathcal{V}}=2 f N_A$ due to an ideal TSS.
Shaded area represents region with $\widetilde{\mathcal{V}}_{2|1}<f N_A$, indicating spin squeezing and entanglement.
Orange dot-dashed curve is a fit to the expected conditional noise reduction with the SNR of the QND measurement as a free parameter.
Inset: Semi-log plot of detected spin squeezing parameter.
Dot-dashed curve is a the expected spin squeezing calculated from the fitted SNR.
Horizontal and vertical error bars represent $1\sigma$ statistical errors, and read-out noise has been subtracted from the data.
\label{fig:squeezing}}
\end{figure}

In Fig.~\ref{fig:squeezing}(a) we plot $\widetilde{\mathcal{V}}_{1,2}(N_A)$, the total measured variance as a function of the number of atoms in the trap for the first two QND measurements (blue circles and green squares).
An ideal TSS has a total variance  $\widetilde{\mathcal{V}}=\langle F^2 \rangle -\langle F\rangle ^2= 2N_A$ (black line in Fig.~\ref{fig:squeezing}(a)). Due to technical noise contribution, the measured variance are higher than the ideal variance for TSS.
The technical noise contribution to $\widetilde{\mathcal{V}}_1$ is indicated by the blue shaded region.
A conditional variance $\widetilde{\mathcal{V}}_{2|1}<f N_A$ (shaded region) indicates spin squeezing and detects entanglement among the atoms~\cite{Toth2007,Toth2009,TothNJP2010,Vitagliano2011a}.
The measured conditional variance $\widetilde{\mathcal{V}}_{2|1}$ (orange diamonds) indicates that we produce spin squeezed states for $N_A>5\times10^5$ atoms.
The  conditional noise for an ideal QND measurement is $\widetilde{\mathcal{V}}_{2|1}=2N_A/(1+\zeta)$, where $\zeta=\tfrac{2}{3}G_1^2N_LN_A$ is the signal-to-noise ratio (SNR) of the measurement~\cite{Hammerer2004,SewellPRL2012}.
A fit to our data (orange dot-dashed line) gives $\widetilde{\mathcal{V}}_{2|1}=2N_A/(1+b \zeta)$ with $b=0.75\pm0.1$, where the reduction in SNR is due to technical noise in the detection system.
In the inset of Fig.~\ref{fig:squeezing}(a) we show the calculated spin squeezing parameter $\xi^2=\widetilde{\mathcal{V}}_{2|1}/f N_A$.
With $N_A=1.1\times10^6$ atoms we measure $\xi^2=0.50\pm0.09$, or 3dB of spin squeezing detected with $5\sigma$ significance.This level of squeezing implies that at least $5.5 \times 10^5$ atoms are entangled with at least one other atom in the ensemble~\cite{TothNJP2010}. While multi-partite entanglement may also be generated in the ensemble~\cite{Urizar-Lanz2014a}, it is not detected by our spin-squeezing inequality~\cite{*[{Multipartite entanglement can also be detected by generalized spin squeezing inequalities. See, for example, Ref. [32] and }] [{}] KorbiczPRL2005}.

We have demonstrated the conditional preparation of a macroscopic singlet state (MSS) via stroboscopic QND measurementin an unpolarized ensemble with more than one million laser-cooled atoms.
We observe 3dB of spin squeezing and detect entanglement with $5\sigma$ statistical significance using a generalized spin squeezing inequality, indicating that at least half the atoms in the sample have formed singlets~\cite{Toth2007,Toth2009,TothNJP2010,Vitagliano2011a}.
Our techniques complement existing experimental methods~\cite{TrotskyPRL2010,SimonN2011,GreifPRL2011,Nascimbene2012a,GreifS2013}, can be readily adapted to measurements of quantum lattice gases~\cite{EckertNP2008,HaukePRA2013} and spinor condensates~\cite{EckertPRL2007}.
In future work we aim to combine our measurement with quantum control techniques~\cite{Behbood2013SC} to produce an unconditionally squeezed macroscopic singlet centered at the origin~\cite{TothNJP2010}, and to use our spatially extended MSS for magnetic field gradiometry~\cite{UrizarPRA2013}.
Due to its SU(2) invariance, the MSS is a good candidate for storing quantum information in a decoherence--free subspace~\cite{Lidar1998a} and for sending information independent of a reference direction~\cite{Bartlett2003a}.


\begin{acknowledgments}
This work was supported by the Spanish MINECO (projects FIS2011-23520 and FIS2012-36673-C03-03), by the EU (projects ERC StG {AQUMET}, ERC  StG GEDENTQOPT and  CHIST-ERA QUASAR), by the Basque Government (Project No. IT4720-10), and by the OTKA (Contract No. K83858).  
\end{acknowledgments}

%


%

\section*{Supplementary material}

\subsection{Atom-light \& atom-field interactions}
We define the collective spin operator ${\bF}\equiv\sum_{i}{\bff}^{(i)}$, where ${\bff}\supi$ is the spin of the $i$'th atom.  
The collective spin obeys commutation relations $[\Fx,\Fy]=i \Fz$.
Probe pulses are described by the Stokes operator ${\bS}$ defined as  $ \hat{S}_i \equiv \tfrac{1}{2} (\hat{a}_+^\dagger,\hat{a}_-^\dagger) \sigma_i (\hat{a}_+,\hat{a}_-)^T $, where the $ \sigma_i $ are the Pauli matrices and $ \hat{a}_\pm $ are annihilation operators for $\sigma_\pm$ polarization, which obey $[\Sx,\Sy]=i \Sz$ and cyclic permutations.  
The input pulses are fully $\Sx$-polarized, i.e. with $\ave{\Sx} = \NL/2$,  $\ave{\Sy}=\ave{\Sz}=0$ and $\Delta^2 S_i = \NL/4$, $i\in\{x,y,z\}$ where $\NL$ is the number of photons in the pulse.

The atoms and light interact via an effective hamiltonian 
\begin{align}
\tau \hat{H}_{\rm eff} =  G_1 \Sz\Fz + G_2 (\Sx\Jx+\Sy\Jy+\hat{S}_0\hat{J}_m/\sqrt{3})
\label{eq:fullHamiltonian}
\end{align}
where $G_1$ and $G_2$ are coupling constants describing vector and tensor lights shifts, respectively, and $\tau$ is the pulse duration~\cite{Echaniz2008,Colangelo2013a}.
The operators $\hat{J}_k\equiv\sum_i^{N_A}\hat{\jmath}_i$ where $\hat{\jmath}_x\equiv\hat{f}_x^2-\hat{f}_y^2$ and $\hat{\jmath}_y\equiv\hat{f}_x\hat{f}_y+\hat{f}_y\hat{f}_x$ describe single-atom Raman coherences, i.e., coherences between states with $m_f$ different by 2, and $\hat{\jmath}_m\equiv(3\hat{f}_z^2-\hat{\mathbf{f}}^2)/\sqrt{3}$ describes the population difference between the $m_f=0$ and $m_f=\pm1$ magnetic sublevels.

The first term in Eq.~(\ref{eq:fullHamiltonian}) describes paramagnetic Faraday rotation: it rotates the polarization in the $\Sx$, $\Sy$ plane by an angle $\phi= G_1 \Fz \ll 1$, and leaves the atomic state unchanged, so that a measurement of $\Sy\supout/\Sx\supin$ indicates $\Fz$ with a shot-noise-limited sensitivity of $\Delta \Fz = \Delta \Sy /G_1$.
Acting alone, this describes a QND measurement of $\Fz$, i.e., with no back-action on $\Fz$. 
The second term, in contrast, leads to an optical rotation $\Sx\rightarrow\Sz$ (due to the birefringence of the atomic sample), and drives a rotation of the atomic spins in the $\Fz$, $\hat{J}_y$ plane (alignment-to-orientation conversion) by an angle $\tan\theta=G_2\Sx/2$~\cite{SewellPRL2012,Colangelo2013a}.
This leads to a detected output
\begin{align}
\Sy\supout=\Sy\supin + G_1\Sx\supin(\Fz\supin+\tan\theta\Jy\supin).
\end{align}
For the experiments described here $\theta\simeq0.3$, and the $\tan\theta$ term can be safely ignored.
The contribution of the remaining terms in Eq.~(\ref{eq:fullHamiltonian}) is negligible.

The atoms interact with the applied magnetic field via the hamiltonian
\begin{align}
\hat{H}_{\rm mag} = - \gamma \bF \cdot \mathbf{B}.
\label{eq:magHamiltonian}
\end{align}
During a single probe-pulse the atomic spins rotate by an angle $\Theta=\gamma B \tau$, where $B=|\mathbf{B}|$.
For our parameters  $\Theta=0.08$ radians, so we can neglect the rotation of the spins during the probe pulses.

\begin{figure}
\centering
\includegraphics[width=\columnwidth]{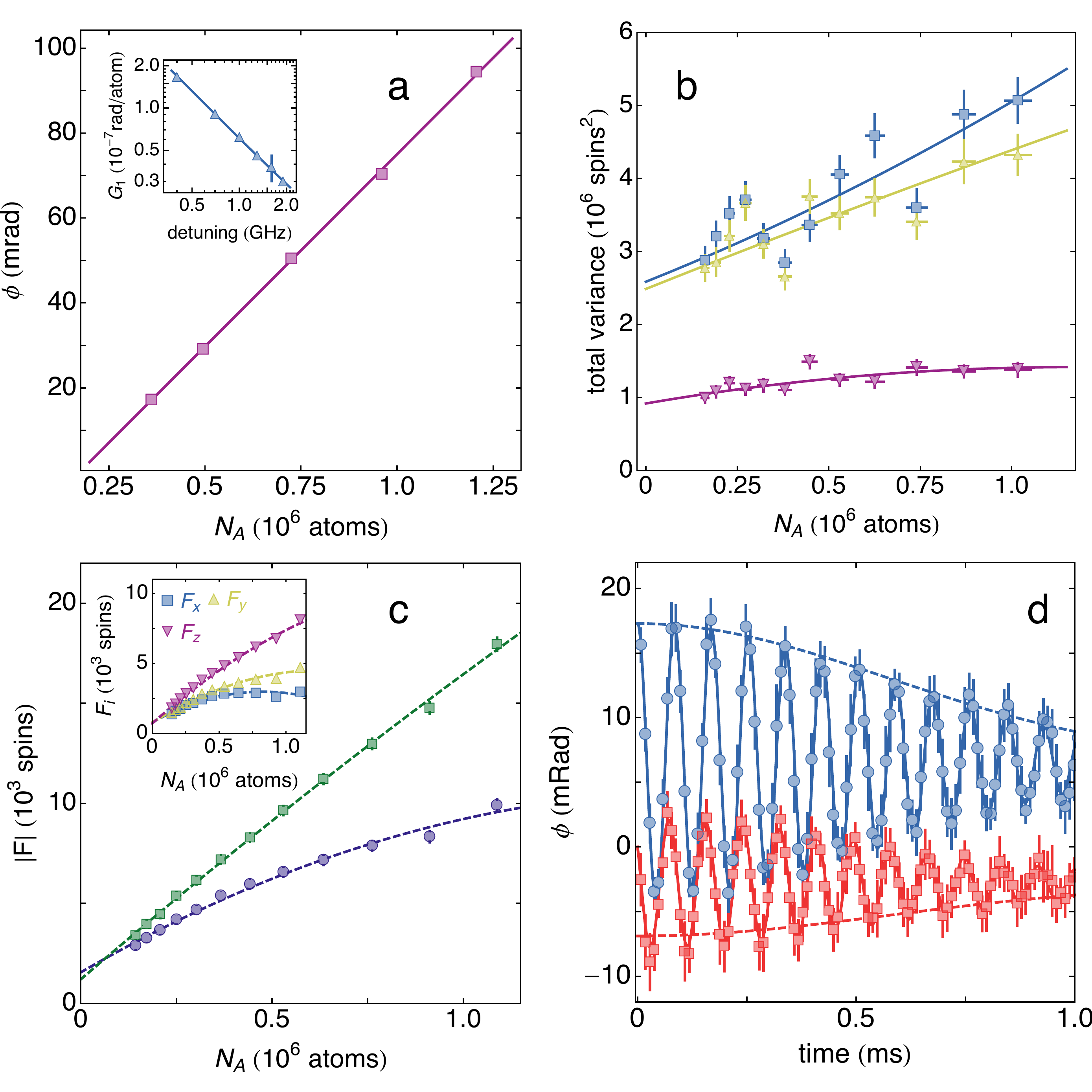}
\caption{(Color online)
(a) Calibration of $G_1$ coupling constant.
We correlate the observed rotation angle $\phi$ against an independent measurement of atom number $N_A$ via absorption imaging.
Inset: from a fit to $G_1$ vs. the probe detuning $\Delta$ we estimate the effective atom-light interaction area $A$ and tensor light shift $G_2$.
(b) Free induction decay (FID) measurement of the applied magnetic field using atoms as an in-situ vector magnetometer.
Blue circles: input $\Fz$-polarized atomic state.
Blue circles: input $\Fy$-polarized atomic state.
Solid line: fit described by Eq.~(\ref{eq:FID}).
Dashed line: gaussian envelope of FID signal.
(c) Length of spin vector $\norm{\hat{F}}$ detected by the first (blue circles) and second (green squares) measurement.
Inset: length of individual spin components $\hat{F}_i$ detected by the first measurement.
(d) Noise scaling of total variance $\mathcal{V}_p=\mathrm{Tr}(\Gamma_p)$ of the first two QND measurements, and conditional variance $\mathcal{V}_{2|1}=\mathrm{Tr}(\Gamma_{2|1})$.
Blue squares: first measurement.
Yellow triangles: second measurement.
Purple inverted triangles: conditional variance.
\label{fig:supp}}
\end{figure}

\subsection{Probe calibration}
The light-atom coupling constant $G_1$ is calibrated by correlating the DANM signal with an independent count of the atom number via absorption imaging~\cite{Kubasik2009,KoschorreckPRL2010a,SewellPRL2012}.
In Fig.~\ref{fig:supp}(a) we show the calibration data.
We find $G_1 = 9.0\pm0.1\times 10^{-8}$ radians per spin at the detuning $\Delta=-700$ MHz.
In the inset of Fig.~\ref{fig:supp}(a) we plot $G_1$ vs. $\Delta$.
We fit this data to find the effective atom-light interaction area $A$~\cite{Kubasik2009}, from which we estimate the tensor light shift $G_2=-4.1_{-0.5}^{+0.4}\times10^{-9}$ radians per spin at $\Delta=-700$ MHz.

\subsection{Noise scaling \& Read-Out Noise}
To estimate the atomic noise contribution to the observed total variance $\mathcal{V}=\mathrm{Tr}(\Gamma)$ of the QND measurements we fit the polynomial $\mathcal{V}(N_A)=\mathcal{V}_0+2 N_A + c N_A^2$ to the measured data, and calculate $\widetilde{\mathcal{V}}_p=\mathcal{V}_p-\mathcal{V}_{0}$, subtracting the read-out noise $\mathcal{V}_{0}$ from the measured total variances.
The data and resulting fits are shown in Fig.~\ref{fig:supp}(b).
The fit to the first (second) measurement yields $\mathcal{V}_0=2.59\pm0.08\times10^6$ ($2.49\pm0.08\times10^6$) and $c=4\pm2\times10^{-7}$ ($1\pm2\times10^{-7}$).
We fit the polynomial $\mathcal{V}_{2|1}(N_A)=\mathcal{V}_{0}+a N_A + c N_A^2$ to the measured conditional variance, giving $\mathcal{V}_{0}=9.2\pm0.8\times10^5$, $a=0.9\pm3$ and $c=-4\pm2\times10^{-7}$, indicating the presence of some correlated technical noise in the detection system.
\begin{figure}
\centering
\includegraphics[width=\columnwidth]{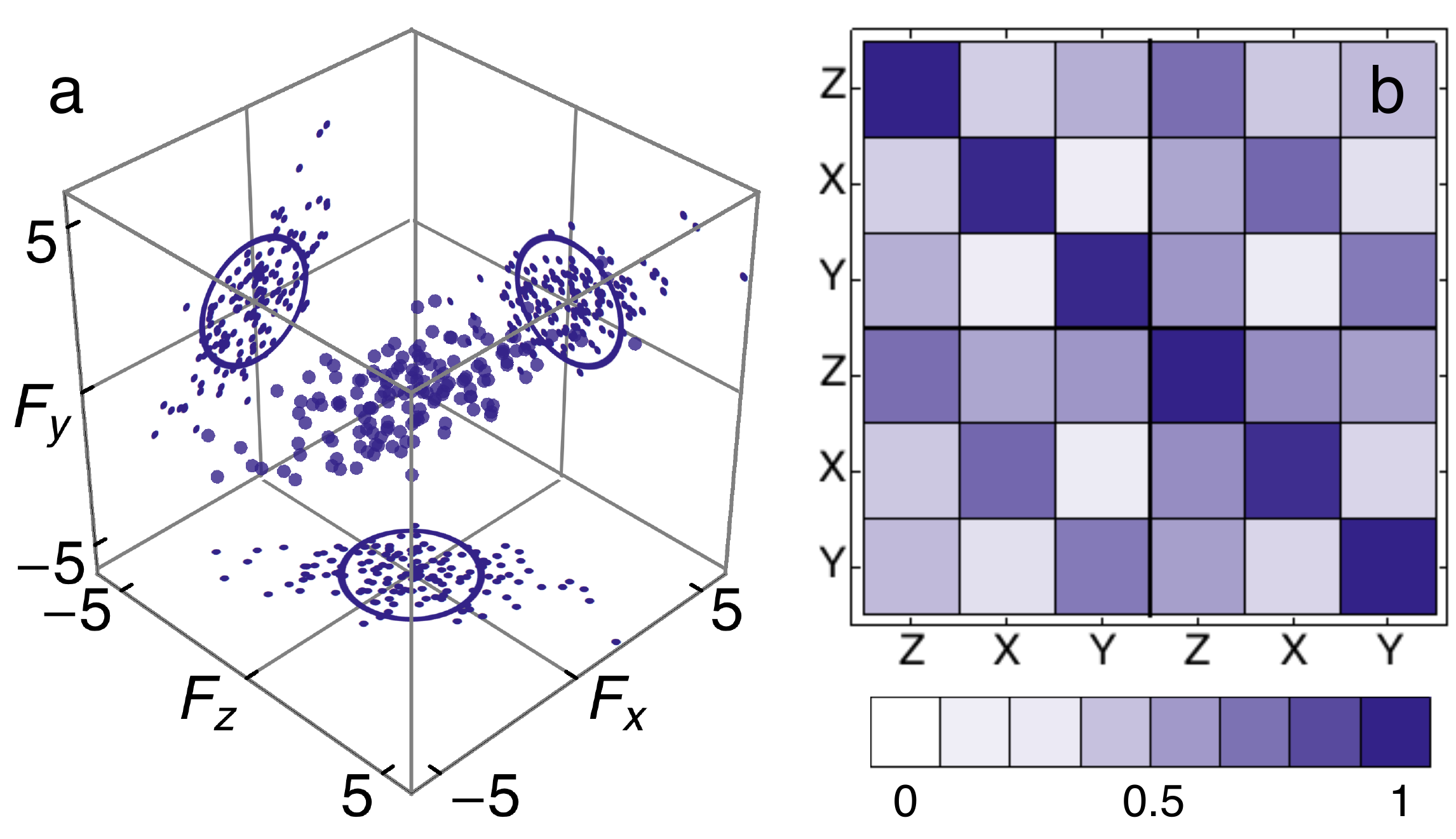}
\caption{(Color online) 
(a) Measured spin distribution (in units of $10^3$ spins) of the input TSS following the state preparation procedure described in the main text.
(b) Correlation matrix between two consecutive three-component collective spin measurements showing strong correlations between measurements of each spin component $\hat{F}_i$.
\label{fig:correlationMatrix}}
\end{figure}

\subsection{Residual polarization}
We observe a small residual atomic polarization due to atoms that are not entangled in the mascroscopic singlet state.
In Fig.~\ref{fig:supp}(c) we plot the length of the spin vector $\norm{\hat{F}}$ detected in the two measurements.
With $N_A=1.1\times10^6$ atoms, we observe a maximum $|F|=13.3\pm0.2\times10^3$ ($18.3\pm0.2\times10^3$) spins for the first (second) measurement, i.e. a residual polarization $\norm{\hat{F}}/(f N_A)=1.66\pm0.02\times10^{-3}$.
In principle with these values we could achieve 20dB of spin squeezing, entangling up to 99\% of the atoms in a macroscopic singlet, before back-action due to the spin uncertainty relations limits the achievable squeezing. 
This residual polarization could be removed by adding a feedback loop to the measurement sequence~\cite{Behbood2013SC}, which would produce an unconditionally squeezed macroscopic singlet centered at the origin.

\subsection{Magnetic field calibration}
We measure the applied magnetic field using the atoms as an in-situ vector magnetometer.
Our technique is described in detail in Ref.~\cite{Behbood2013VB}.
We polarize the atoms via optically pumping along first $\Fz$ and then $\Fy$, and observe the free induction decay (FID) of the resulting Larmor precession using the Faraday rotation probe.
We model density distribution along the length of the trap with a gaussian $A \exp(-(z-z_0)^2/2\sigma^2)$, with an RMS width $\sigma= 2.68\pm0.3$~mm.
A typical density profile and gaussian fit is shown in Fig.~\ref{fig:supp}(d).
This leads to observed signals for the two input states
\begin{widetext}
\begin{align}
	\theta(t)&=\frac{G_1}{B^2}
	\begin{cases}
		\left( B_z^2 +\left(B_x^2+B_y^2\right)\cos\omega\exp\left(-t^2/T_2^2\right) \right) F_z(0) \\
		\left( B_yB_z\left(1-\cos\omega\exp\left(-t^2/T_2^2\right) \right) + B_x B \sin\omega \exp\left(-t^2/T_2^2 \right) \right) F_y(0)
	\end{cases}
\label{eq:FID}
\end{align}
\end{widetext}
where $\omega=\gamma B t$, $B=|\mathbf{B}|$, and $\gamma=\mu_B g_f/\hbar$ is the atomic gyromagnetic ratio.
By fitting theses signals, we extract the vector field $\mathbf{B}$ and the FID transverse relaxation time $T_2=1/(\sigma \gamma \partial B/\partial z)$.
For the data shown we find $B_x=9.6\pm0.4$ mG, $B_y=9.7\pm0.4$ mG, $B_z=9.9\pm0.1$ mG and $T_2=745\pm45$ $\mu$s.

\subsection{Input state}
In Fig.~\ref{fig:correlationMatrix}(a) we plot the spin distribution ${\bf F}^{(1)}$ of the collective spin of a sample with $N_A=1.4\times10^6$ atoms measured by the first three probe pulses.
We measure an initial spin covariance matrix of
\begin{align}
\Gamma_1&=
\left(
\begin{array}{ccc}
 1.90 & 1.10 & 1.10 \\
 1.10 & 1.40 & 0.81 \\
 1.10 & 0.81 & 1.30 \\
\end{array}
\right)
\times 10^6 \mathrm{\, spins^2}.
\end{align} 
For comparison, an ideal TSS would have $\Gamma=\mathrm{diag}(0.93,0.93,0.93)\times10^6$ spins$^2$ with the same number of atoms.
The larger measured variances, and non-zero covariances, in $\Gamma_1$ indicate the presence of atomic technical noise due to imperfect state preparation and shot-to-shot fluctuations in the atom number and applied magnetic field.

\subsection{Measurement correlations}
In Fig.~\ref{fig:correlationMatrix}(b) we plot the correlations $\rho_{ij}\equiv\cov(\hat{F}_i,\hat{F}_j)/\Delta\hat{F}_i\Delta\hat{F}_j$ between the first six QND measurements.
The off-diagonal elements indicate that successive measurements of the same spin component $\hat{F}_k$ are well correlated.
This allows us to predict the outcome of the second measurements $F_k^{(2)}$ with a reduced conditional uncertainty.
The residual correlation between measurements of different spin components is due to correlated technical noise in the atomic state preparation, and in the detection system.

\end{document}